\documentclass[10pt,twocolumn]{article}

\usepackage[margin=1.9cm]{geometry}
\usepackage[T1]{fontenc}
\usepackage[utf8]{inputenc}
\usepackage{amsmath,amssymb}
\usepackage{booktabs}
\usepackage{graphicx}
\usepackage[expansion=false]{microtype}
\usepackage{url}
\usepackage[hidelinks]{hyperref}
\usepackage{caption}
\usepackage[compact]{titlesec}
\titlespacing{\section}{0pt}{1.2ex plus .2ex}{0.6ex plus .1ex}
\titlespacing{\subsection}{0pt}{1.0ex plus .2ex}{0.4ex plus .1ex}

\title{\vspace{-1.2em}\bfseries Reviewer Capability Governs Rejection
Targeting, Not Repair Skill: Evidence from LLM Execute--Review--Revise
Pipelines\vspace{-0.4em}}

\author{Faizan Tanveer\\[2pt]
\small National University of Computer and Emerging Sciences (FAST NUCES)\\
\small \texttt{l230640@lhr.nu.edu.pk}}
\date{}

\begin{document}
\maketitle

\begin{abstract}
\noindent
Multi-agent LLM pipelines increasingly assign roles, including execution and
verification, to models of different capability tiers. This is done because
running a flagship model at every stage is expensive. Previous literature has
established that verification stages are not always beneficial, but holds
reviewer capability roughly fixed relative to the executor. We vary it. We
replace the reviewer with models spanning a capability range down to one that
cannot solve the problems at all, and measure the outcome of every individual
rejection. This is done across a constant set of 100 olympiad mathematics
problems.

A cross-family mid-tier reviewer improves final accuracy by 12 percentage
points, from 52 to 64 percent ($p = 0.0005$), with zero damaged answers.
Same-model self-review attains the highest error-detection rate of any condition
(0.85 recall) yet yields no significant gain: it rejects 2.1 times as often for
a third the repair rate (15 against 43 percent, $p = 0.0074$) and falsely
rejects 35 percent of its own correct answers against 2 percent for the
cross-family reviewer (paired $p = 0.000015$). The low damage rate of
self-review proves to be an artifact of revision inertia rather than reviewer
quality: of 18 falsely rejected correct answers, the three where the executor
complied all became wrong, while the fifteen it ignored survived unchanged.
Below a capability floor the role becomes inert: our weakest reviewer changed
zero of 100 final answers while doubling token cost. These findings describe a single executor--reviewer configuration on 100 problems and should be read as a controlled pilot rather than a general claim about verification stages.
\end{abstract}

\section{Introduction}

Multi-agent LLM pipelines increasingly assign planning, execution, and
verification to separate model instances, often using cheaper models for less
demanding roles. This raises a basic question: which roles can be delegated to
weaker models without degrading outcomes?

Recent work suggests that verification is not reliably beneficial. Detection
Without Correction \cite{nilayam2026detection} decomposes downstream responses
into error detection and conditional revision, finding that when an error is
flagged, the resulting revision is wrong 53 to 94 percent of the time across
four model families and four benchmarks. Yang et al.\ \cite{yang2026precise}
further show that reviewer precision and critique uptake are distinct: their
planner--executor--reviewer pipeline achieved higher reviewer precision than
broadcast peer discussion (0.861 against 0.644) but performed worse because its
critiques were less likely to affect the final answer. AgentCARD
\cite{jiang2026specialize} shows that heterogeneous teams can outperform
homogeneous teams at equivalent cost.

These studies largely hold reviewer capability fixed relative to the executor.
We vary it. Holding a capable executor constant across 100 olympiad mathematics
problems, we replace the reviewer with models spanning a wide capability range,
including a model unable to solve the problems itself. We measure not only final
accuracy but the outcome of every rejection: whether it repaired an incorrect
answer, damaged a correct one, or changed nothing.

The results challenge the assumption that better detection produces better
outcomes. A cross-family mid-tier reviewer improves accuracy by 12 percentage
points with zero damaged answers. Same-model self-review achieves the highest
detection rate of any condition but produces no significant accuracy gain: it
rejects 2.1 times as often while repairing only one-third as many answers, and
falsely rejects 35 percent of its own correct answers. Its apparently low damage
rate is misleading. Among false rejections, every case in which the executor
complied with the critique produced a wrong answer; the remaining cases were
harmless only because the executor ignored the critique. Thus, revision
conservatism can mask reviewer incompetence.

At the bottom of the capability range, the reviewer becomes effectively inert.
Our weakest reviewer changes none of the 100 final answers while doubling token
cost.

\paragraph{Contributions.}
(i)~We introduce a rejection-outcome decomposition that separates repair,
damage, ignored critique, and miscorrection across reviewer capability tiers.
(ii)~We show that the failure of detection without correction depends on
reviewer capability rather than being a uniform property of verification.
(iii)~We show that the apparent safety of self-review can arise from revision
inertia rather than reviewer quality. (iv)~We identify a capability floor below
which review produces no measurable change at non-trivial cost.

\section{Related Work}

\paragraph{Role assignment and cost.}
AgentCARD \cite{jiang2026specialize} benchmarks LLM agent teams across role
assignments and deployment modes, finding that heterogeneous teams occupy the
cost--accuracy Pareto frontier: they achieve up to 44 percent higher accuracy
than cost-equivalent homogeneous teams, or comparable accuracy at roughly
$12\times$ lower cost. It also finds that role bottlenecks are domain-dependent.
AgentCARD treats roles as slots to assign, however, rather than reviewer
capability as a variable to sweep.

\paragraph{Reviewer precision versus critique uptake.}
Yang et al.\ \cite{yang2026precise} evaluate hierarchical reviewer designs on
4{,}181 verifier-grounded Omni-MATH problems using matched actors. Their
planner--executor--reviewer pipeline has a more precise reviewer than broadcast
peer discussion (0.861 against 0.644) yet performs worse because useful
critiques are less likely to change the candidate carried forward. Thus,
detection quality and critique uptake are empirically separable. Their reviewers
remain at the same capability tier as the other actors.

\paragraph{Verification harm as a mechanism.}
Detection Without Correction \cite{nilayam2026detection} decomposes downstream
response into detection and conditional generation, identifying
detect--miscorrect as a key failure mode. Conditional miscorrection rates reach
53 to 94 percent across four model families, four benchmarks, and two methods.
However, this mechanism is established using near-peer models; reviewer
capability is not varied systematically.\paragraph{Self-correction and mistake finding.} A parallel literature studies review by the generating model itself. Self-Refine \cite{madaan2023selfrefine} reports gains from iterative self-feedback, while Huang et al. \cite{huang2024selfcorrect} find that intrinsic self-correction does not improve reasoning and often degrades it. Tyen et al. \cite{tyen2024errors} locate the cause in detection rather than correction: models struggle to find logical mistakes, but correct them reliably when the error location is supplied. Our self-review condition inverts that account. It attains the highest recall of any condition we test (0.85) and still repairs only 15 percent of what it rejects, so poor outcomes here follow from imprecise targeting rather than from failure to detect. Reviewer self-preference is documented for LLM judges \cite{zheng2023judging}; we observe the opposite sign, with the reviewer falsely rejecting 35 percent of its own correct answers. Multi-agent debate \cite{du2024multiagent} improves accuracy through cross-model critique, which our capability floor bounds from below: below some reviewer competence the exchange produces no change at all.

\paragraph{This work.}
All four lines of work hold reviewer capability approximately fixed relative to
the executor. We instead hold the executor fixed and sweep reviewer capability,
including a reviewer that solves approximately none of the problems unaided. We
then decompose every rejection into its outcome: repair, damage, ignored
critique, or miscorrection, instead of evaluating review only through aggregate
final accuracy.

\section{Method}

\subsection{Pipeline}

The pipeline has two stages and no planner: an executor produces a solution, a
reviewer accepts or rejects it, and on rejection the executor revises once. The
planner stage is omitted deliberately. The question here concerns the reviewer,
and an additional stage would add cost and make it harder to attribute an
outcome to the reviewer alone.

For each problem the sequence is as follows. The executor produces a solution
$A_0$. The reviewer sees the problem and $A_0$, but never the ground truth, and
emits either \textsc{accept} or \textsc{reject} with a written critique. If the
verdict is \textsc{reject}, the executor receives its own solution together with
the critique and produces a single revision $A_1$. The final answer is $A_1$
when the solution was rejected and $A_0$ otherwise.

The design is paired. The executor solves each problem once, and the identical
solution $A_0$ is then reviewed under every condition. Any difference between
conditions is therefore attributable to the reviewer rather than to executor
sampling variance, and every comparison between conditions is paired at the
problem level.

Reviews that are empty or cannot be parsed are treated as \textsc{accept} and
flagged. The rate at which this occurs is reported for each condition.

\subsection{Conditions}

The executor is fixed across all conditions: Gemini 3.1 Flash-Lite, which solves
52 percent of this task set unaided. Four conditions vary the reviewer. C0 is no
review, in which $A_0$ is final. C1 is self-review, in which the reviewer is the
same model as the executor. C2 is a cross-family mid-tier reviewer,
gpt-oss-20b. C3 is a weak reviewer, Llama-3.1-8B, which solves approximately
none of these problems unaided.

The three models come from three different families, which allows self-review to
be distinguished from cross-family review rather than confounding capability
with model identity. Temperature is 0.7 for every call. Per-model output token
limits are recorded in the repository.

\subsection{Tasks}

Problems are drawn from Omni-MATH \cite{gao2024omnimath}, restricted to
difficulty band 5.0 to 6.0. Three filters are applied. First, only problems
whose reference answer is a plain integer or decimal are retained, which
excludes 817 problems in the band. Second, malformed problem types are excluded,
specifically problems that reference another problem, open-ended estimation
problems, and guts-round chains, removing a further 28. Third, 2{,}889 problems
fall outside the difficulty band. The remaining problems are shuffled with a
recorded seed and the first 100 are used.

The restriction to numeric answers keeps grading exact and fully automated, with
no language model acting as judge anywhere in the scoring path. It is also a
selection bias that should be stated plainly: it favours problems from number
theory and combinatorics and excludes problems whose answers are proofs or
symbolic expressions.

\subsection{Grading}

An answer is correct when it matches the reference value exactly after
canonicalisation. The answer is read from a line of the form
\texttt{ANSWER: <value>}, with documented fallbacks for cases where a model does
not comply: the last numeric value in the response, evaluation of a fraction
such as $31/5$, stripping of \LaTeX{} delimiters, and rejection of non-finite
values. The same grader scores $A_0$ and the final answer, and no human or model
judgment enters the scoring path.

\subsection{Reviewer input}

An initial run revealed that the mid-tier reviewer's reviews were being
truncated by its provider's 8{,}000 tokens-per-minute rate limit on 29 percent of problems, returning
empty text that was then counted as \textsc{accept}. Because the rate limit
is fixed, the only available remedy was to reduce reviewer input. Two changes
were made. The answer-format instructions intended for the executor were
stripped from the problem text shown to the reviewer, and the solution was
truncated to its final 2{,}000 characters with an explicit marker indicating
that earlier steps had been omitted. Together these reduce reviewer input by
approximately 66 percent on long problems.

The change was applied identically to all reviewer conditions and every review
was regenerated. Partial results were available at $n = 18$ and $n = 36$ before this change was made. The change was triggered by the reviewer's malformed-output rate reaching 29 percent, not by inspection of accuracy, and the results reported here are exclusively from the post-change run at $N = 100$; pre-change figures are not pooled or reported. After the change the rate of unparseable reviews was 0 percent
for C1, 16 percent for C2, and 0 percent for C3.

One side effect is worth recording. Input length affects what the weak reviewer
produces. Under the original prompt it returned mostly empty text; under the
shortened prompt it produces fluent, hedged prose while still accepting almost
everything. The C3 results therefore describe the weak reviewer under the
shortened prompt and should not be read as a property of the model in general.

\subsection{Metrics}

For each condition we report final accuracy, recall defined as the probability
of rejection given that $A_0$ was wrong, and the false-rejection rate defined as
the probability of rejection given that $A_0$ was correct.

Aggregate accuracy conceals what a rejection actually accomplishes, so every
rejection is additionally classified into one of four outcomes. A \emph{repair}
changes a wrong answer into a correct one. A \emph{damage} changes a correct
answer into a wrong one. A \emph{no-change} means the revision returned the
identical answer, and is reported separately according to whether $A_0$ was
correct, because a no-change on a correct answer is a false rejection that
happened to be harmless. A \emph{miscorrection} changes a wrong answer into a
different wrong answer.

Token cost is taken from exact provider usage metadata rather than estimated,
and is reported both per task and as the number of additional tokens spent per
additional correct answer relative to C0.

Recall should be read alongside rejection volume. A reviewer that rejects
indiscriminately obtains high recall cheaply, and this turns out to matter here.

\subsection{Statistics}

Accuracy comparisons use exact McNemar tests on the paired per-problem outcomes,
which is appropriate because the same problems appear in every condition and
only discordant pairs carry information. Comparisons of proportions computed
over rejections use Fisher's exact test. Conditions with fewer than five
rejections are described but not tested, since a test on so few observations
would not be meaningful. Pairwise comparisons are reported uncorrected for multiplicity. At this pilot scale the large effects would survive standard corrections, but the borderline comparisons should be read as weaker than their nominal $p$-values suggest.

\section{Results}

\subsection{Final accuracy}

\begin{table}[t]
\centering\small
\caption{Final accuracy by condition. Exact McNemar on paired per-problem
outcomes ($N = 100$).}
\label{tab:accuracy}
\begin{tabular}{@{}llll@{}}
\toprule
Condition & Accuracy & vs.\ C0 & McNemar $p$ \\
\midrule
C0, no review     & 0.52 & --- & --- \\
C1, self-review   & 0.58 & $+6$ pp  & 0.146, ns \\
C2, cross-family  & 0.64 & $+12$ pp & \textbf{0.0005} \\
C3, weak          & 0.52 & 0        & 1.000 \\
\bottomrule
\end{tabular}
\end{table}

The cross-family reviewer improves accuracy by 12 percentage points and the
improvement is entirely one-directional. There are 12 problems that C2 answered
correctly and C0 did not, and none in the reverse direction.

C3 produced no discordant pairs at all. Across 100 problems it did not change a
single final answer, so its accuracy is identical to the no-review baseline by
construction rather than by coincidence. This inertness was measured only under the shortened reviewer prompt described in Section 3.5; the same model behaved differently under the original prompt, so the capability floor reported here is prompt-conditional rather than a fixed property of the model.

Self-review improves accuracy by 6 percentage points, which does not reach
significance at this sample size. We therefore do not claim that self-review
improves accuracy, and we do not claim that the cross-family reviewer beats
self-review on accuracy either, a comparison that is directionally consistent
but underpowered here ($p = 0.238$).

\subsection{Detection does not predict benefit}

\begin{table}[t]
\centering\small
\caption{Detection behaviour. Recall is $P(\textsc{reject} \mid A_0$ wrong$)$;
the false-rejection rate is $P(\textsc{reject} \mid A_0$ correct$)$, over the 52
problems where $A_0$ was correct.}
\label{tab:detection}
\begin{tabular}{@{}llll@{}}
\toprule
Condition & Recall & False-rej.\ rate & Rejections \\
\midrule
C1, self-review  & 0.85 & 18/52, 35\% & 59 \\
C2, cross-family & 0.56 & 1/52, 2\%   & 28 \\
C3, weak         & 0.04 & 0/52, 0\%   & 2  \\
\bottomrule
\end{tabular}
\end{table}

Self-review detects errors at the highest rate of any condition and yet produces
no significant accuracy gain, while the cross-family reviewer detects fewer
errors and produces the only significant gain. The explanation is visible in the
third column. Self-review achieves its recall by rejecting frequently, including
35 percent of answers that were already correct.

\subsection{Rejection outcomes}

\begin{table*}[t]
\centering\small
\caption{Every rejection classified by outcome. No-change is split by whether
$A_0$ was correct, because a no-change on a correct answer is a false rejection
that happened to be harmless. Percentages are of that condition's rejections.}
\label{tab:outcomes}
\begin{tabular}{@{}lllllll@{}}
\toprule
Condition & Rejections & Repair & Damage & No-change ($A_0$ wrong) & No-change ($A_0$ right) & Miscorrection \\
\midrule
C1 & 59 & 9, 15\%  & 3, 5\% & 9, 15\%  & 15, 25\% & 23, 39\% \\
C2 & 28 & 12, 43\% & 0, 0\% & 3, 11\%  & 1, 4\%   & 12, 43\% \\
C3 & 2  & 0        & 0      & 2        & 0        & 0        \\
\bottomrule
\end{tabular}
\end{table*}

Comparing C1 and C2 over their rejections, the repair rate differs, 15 percent
against 43 percent, Fisher exact $p = 0.0074$, and so does the rate at which a
rejection produces no change at all, 41 percent against 14 percent,
$p = 0.0151$.

Expressed as selectivity, self-review requires 6.6 rejections for each useful
repair while the cross-family reviewer requires 2.3.

\subsection{Revision inertia conceals reviewer error}

Of C1's 18 false rejections (correct answers rejected), the executor changed its
answer in 3 cases, and all 3 became wrong; in the remaining 15, the revision
returned the identical correct answer. Self-review's low damage rate is
therefore not evidence of reviewer quality: it is largely a consequence of the
executor ignoring 83 percent of its false rejections. As a bounded counterfactual, if
the executor had complied with every one of C1's false rejections, its accuracy
would have fallen below the no-review baseline. This does not predict that the
executor would in fact have complied; it isolates the damage that the reviewer's
false rejections would have caused under full uptake.

\subsection{Targeting and repair skill are separable}

\begin{table}[t]
\centering\small
\caption{Targeting versus repair skill. Repair skill conditions on the executor
having actually changed the answer, excluding no-change cases.}
\label{tab:targeting}
\begin{tabular}{@{}lll@{}}
\toprule
 & False-rej.\ rate & Repair $\mid$ change \\
 & (targeting) & (skill) \\
\midrule
C1, self-review  & 18/52, 35\% & 9/35, 26\% \\
C2, cross-family & 1/52, 2\%   & 12/24, 50\% \\
\midrule
Test & McNemar & Fisher \\
$p$  & \textbf{0.000015} & 0.096 \\
\bottomrule
\end{tabular}
\end{table}

The two columns behave differently. Targeting separates the conditions very
sharply: over the 52 problems where $A_0$ was correct, there are 17 problems
that self-review falsely rejected and the cross-family reviewer did not, and
zero in the reverse direction; the cross-family reviewer's single false rejection was on a problem self-review also rejected.

Repair skill does not separate the conditions at this sample size, 26 percent
against 50 percent, $p = 0.096$, though the direction is consistent with the
cross-family reviewer repairing more effectively. With 35 and 24 attempted
changes respectively this comparison is underpowered, and we report an absence
of evidence rather than evidence of absence. The miscorrection rates are close
as well, 39 percent against 43 percent.

Reviewer capability therefore governs which answers get rejected far more
clearly than it governs how well a rejection is subsequently repaired.

\subsection{Cost}

\begin{table}[t]
\centering\small
\caption{Token cost. Counts are exact provider usage metadata, not estimates.}
\label{tab:cost}
\begin{tabular}{@{}llll@{}}
\toprule
Cond. & Tokens/task & $\Delta$ tokens & Tokens per extra correct \\
\midrule
C0 & 668   & ---     & --- \\
C1 & 2{,}207 & $+1{,}539$ & $\approx 25{,}600$ \\
C2 & 3{,}511 & $+2{,}844$ & $\approx 23{,}700$ \\
C3 & 1{,}353 & $+685$     & undefined \\
\bottomrule
\end{tabular}
\end{table}

The cross-family reviewer is not cheaper per additional correct answer than
self-review; the two are close, at roughly 24{,}000 and 26{,}000 tokens
respectively. What it buys at that price is twice the accuracy improvement with
no damaged answers. The weak reviewer consumes roughly twice the baseline token
cost and returns nothing, so its cost per additional correct answer is
undefined. This is the clearest form of the capability floor: below some level of
reviewer competence, the role is not merely less useful but strictly wasteful.

\section{Discussion}

Our results clarify the distinction between reviewer detection and reviewer
utility. Yang et al.\ \cite{yang2026precise} show that these two can differ: a
reviewer can identify errors more reliably without producing a better downstream
answer. Across capability tiers, however, we find a stronger pattern. Detection
and utility are not merely separable; they are inversely ordered. The reviewers
that reject the most are not the reviewers that produce the largest gains.
Same-model self-review achieves the highest detection rate but does not improve
final accuracy, while a cheaper cross-family mid-tier reviewer produces the
largest gain while making no observed damaging changes.

The results also reproduce the detect-without-correct phenomenon
\cite{nilayam2026detection} while showing that its expression depends on
reviewer capability, though not in the way aggregate miscorrection rates would
suggest. What separates the conditions is how often a critique is generated and
then ignored entirely: 41 percent of self-review rejections left the answer
unchanged, against 14 percent for the cross-family reviewer. Miscorrection given
an attempted change is almost identical across the two, 39 percent against 43
percent. This matters because detect-without-correct is often discussed as a
property of verification itself. Our results suggest instead that what varies
with reviewer capability is which answers are rejected and whether the critique
is acted upon at all, rather than the quality of the correction once a change is
attempted.

The self-review results expose a further problem with aggregate damage rates. C1
has a low observed damage rate, but this is not because its critiques are
reliably correct. Of its 18 false rejections, the executor acted on only 3, and
all 3 revisions became wrong; the other 15 were effectively harmless because the
executor returned the original correct answer. Low observed damage can therefore
be produced by revision inertia: a poor reviewer generates harmful critiques,
but a conservative executor prevents those critiques from reaching the final
answer. Reporting low damage without measuring critique uptake can make an
incompetent reviewer appear safe. The bounded counterfactual makes the point
directly: had the executor complied with all 18 false rejections, C1's accuracy
would have fallen below the no-review baseline. This isolates the damage latent
in the reviewer's false rejections.

These findings suggest a practical criterion for reviewer selection. Reviewer
precision should matter more than detection rate. A reviewer that flags fewer
errors but is usually right when it intervenes can be more valuable than one
that detects more errors but generates unreliable corrections. In our
experiments the cross-family reviewer matched or exceeded same-model
self-review on every rejection-outcome measure we recorded, including repair rate, damage rate, and targeting. The final-accuracy comparison between these
two conditions is directionally consistent but did not reach significance at this sample size ($p = 0.238$), and we do not claim it. This provides a concrete reason not to select
reviewers simply by asking which model is best at finding mistakes.

\subsection{Qualitative failure modes}

The transcripts reveal three recurring failure modes that complement the
quantitative decomposition. These three examples were selected for the clarity with which they exhibit each failure mode rather than by systematic sampling, and should be read as illustrations of the mechanisms rather than as evidence of their frequency. First, C1 produced fabricated
justifications: it identified a plausible-looking issue and supplied a
mathematical rationale that did not actually follow from the executor's work.
Second, it produced confidently wrong critiques, where the reviewer asserted
that a correct derivation contained an error and proposed a revision that moved
the solution away from the correct answer. Third, the weakest reviewer often
produced fluent but vacuous endorsements: rather than independently checking the
argument, it restated the executor's reasoning in polished language and declared
it correct.

For example, C1 rejected a correct answer to \texttt{omni\_0076} on the grounds
that ``the reasoning is flawed because it assumes that for any general triangle,
the 17 non-collinear subsets of 3 points will always form 17 distinct circles,''
invoking the nine-point circle to argue that the count must be smaller; the
executor complied and revised 17, the reference answer, to 7. In
\texttt{omni\_0083} the same reviewer asserted that ``the maximum number of
bishops that can be placed on an $8\times 8$ board such that at most $k$ bishops
lie on any diagonal is given by the formula $2(n-1)k$,'' a formula the reviewer
does not justify, and which yields 42 where the reference answer is 38. By
contrast, the weakest reviewer endorsed an incorrect solution to
\texttt{omni\_0002} with ``None. The solution correctly identifies the basin of
attraction of the origin as the set $S$, and the boundary of $S$ as the circle
$x^2 + (y-1/2)^2 = 1/4$. The area of this circle is correctly calculated as
$\pi/4$, which leads to the correct value of $r$ and the final answer of 6,''
where the reference answer is 133.

These examples illustrate why surface-level critique quality can be misleading.
A response can be articulate, specific, and mathematically formatted while
contributing no genuine verification. The quantitative rejection decomposition
captures the downstream consequence of these behaviours; the transcripts show
how they arise.

\section{Limitations}

The sample is 100 problems from a single benchmark, in one difficulty band, with
one executor, one revision round, and one reviewer prompt template. The findings
should be read as describing this configuration rather than review stages in
general.

The comparison between the cross-family reviewer and self-review on final
accuracy is directionally consistent but underpowered at this sample size
($p = 0.238$). We do not claim it. The claims we do make about the difference
between these conditions concern rejection behaviour, where the denominators are
the rejections themselves and the effects are large enough to detect.

The recall figure for the cross-family reviewer is a lower bound. Sixteen
percent of its reviews terminated at the provider's 8{,}000 tokens-per-minute rate limit and were
conservatively counted as accepts, which can only reduce its measured recall.

Several proportions rest on small denominators, in particular the 3 damage cases
for self-review and the repair counts of 9 and 12. Denominators are shown
throughout so that the reader can judge the weight of each figure directly.

The restriction to problems with numeric answers selects a non-random subset of
the benchmark, favouring number theory and combinatorics and excluding proof and
expression-valued problems.

The C3 results describe the weak reviewer under the shortened reviewer prompt.
Under the original longer prompt the same model returned mostly empty text, so
its behaviour is sensitive to input length in a way that limits how far the
inertness result generalises.

Free-tier provider limits constrained both the choice of models and the sample
size.

\section{Reproducibility}

The analysis plan was committed before the experiment was run, and the original
plan is retained in the repository unchanged. It describes a different study: whether cross-model error diversity predicts multi-agent benefit. Calibration probes falsified that design's premise for the available models. Across four generated task families, including the hardest configurations the generators could produce, two of the three models answered essentially every item correctly, so the error-diversity metric could not be computed; moving to olympiad problems produced errors from the strong model but none at all from the weak one. No difficulty band existed at which all three models erred comparably. The study was therefore re-scoped to treat the capability spread itself as the object of study, and the amendment documenting this was written before the reported experiment was run. The pipeline mechanics, paired design, exact-match numeric grading, models, and temperature are carried over unchanged. Both documents are retained as a record of the project's methodological trajectory. The amendment also records the reviewer-input change described in Section 3.5, which followed
calibration. All prompts are frozen and stored in the repository. The harness
caches every model call to disk and resumes cleanly after interruption, so the
run is reproducible without re-spending API quota. Exact model identifiers,
sampling temperature, seeds, and per-call token usage are recorded. Code and
logs are available at
\url{https://github.com/faizan-tnvr004/llm-reviewer-capability}.

\bibliographystyle{plain}
\bibliography{references}

\end{document}